\documentclass[aps,prb,showpacs,twocolumn]{revtex4-1}
\usepackage{amsfonts}
\usepackage{amssymb}
\usepackage{amsmath}
\usepackage{graphicx}
\usepackage{epsfig}

\begin{document}

\title{Topological phases in Kitaev chain with imbalanced pairing}
\author{C. Li${}^1$, X. Z. Zhang${}^2$, G. Zhang${}^2$ and Z. Song${}^1$}
\email{songtc@nankai.edu.cn}
\affiliation{${}^1$School of Physics, Nankai University, Tianjin 300071, China \\
${}^2$College of Physics and Materials Science, Tianjin Normal University,
Tianjin 300387, China}

\begin{abstract}
We systematically study a Kitaev chain with imbalanced pair creation and
annihilation, which is introduced by non-Hermitian pairing terms. Exact
phase diagram shows that the topological phase is still robust under the
influence of the conditional imbalance. The gapped phases are characterized
by a topological invariant, the extended Zak phase, which is defined by the
biorthonormal inner product. Such phases are destroyed at the points where
the coalescence of groundstates occur, associating with the time-reversal
symmetry breaking. We find that the Majorana edge modes also exist for the
open chain within unbroken time-reversal symmetric region, demonstrating the
bulk-edge correspondence in such a non-Hermitian system.
\end{abstract}

\maketitle



\section{Introduction}

The Kitaev model is employed to describe the dynamics of spinless fermions
with superconducting $p$-wave pairing \cite{Kitaev}. The topological
superconducting is demonstrated by unpaired Majorana modes exponentially
localized at the ends of open Kitaev chains \cite{Sarma,Stern}. Intimately
related to the superconducting phase,\ the pairing term takes the key role
in the Kitaev model, which violates the conservation of fermion number but
preserves the parity of the number. It is the fermionized version of the
familiar one-dimensional transverse field Ising model \cite{Pfeuty}, which
is one of the simplest solvable models exhibiting quantum criticality and
demonstrating a spontaneous symmetry breaking driven quantum phase
transition \cite{SachdevBook}.\textbf{\ }So far, most of the investigations
on this model focus on the case with the equal magnitude of amplitudes for
pair creation and annihilation, i.e., the balanced kitaev model\textbf{. }%
Intuitively, the balance between pair production and annihilation takes an
important role in the existence of the gapped superconducting phase. When
the amplitude norm of the pair production is not equal to that of
annihilation, the pair state may be destroyed. A natural question is what
happens when the balance is broken. Theoretically, the imbalanced pair terms
will break the Hermiticity of the Hamiltonian. Nowadays, non-Hermitian
Hamiltonian is no longer a forbidden regime in quantum mechanics since the
discovery that a certain class of non-Hermitian Hamiltonians could exhibit
the entirely real spectra \cite{Bender,Bender1}.\textbf{\ }The aim of this
paper is to answer this question based on the non-Hermitian quantum theory.

The origin of the reality of the spectrum of a non-Hermitian Hamiltonian is
the pseudo-Hermiticity of the Hamiltonian operator \cite{Ali1}. Such kinds
of Hamiltonians possess a particular symmetry, i.e., it commutes with the
combined operator $PT$, but not necessarily with $P$\ and $T$\ separately.
Here $P$\ is a unitary operator, such as parity, translation, rotation
operators etc., while $T$\ is an anti-unitary operator, such as
time-reversal operator. The combined symmetry is said to be unbroken if
every eigenstate of the Hamiltonian is $PT$-symmetric; then, the entire
spectrum is real, while is said to be spontaneously broken if some
eigenstates of the Hamiltonian are not the eigenstates of the combined
symmetry. The study of non-Hermitian Kitaev model \cite%
{Law,Tong,Yuce,You,Klett,Menke} and Ising model \cite{ZXZ,LC} has performed
within the pseudo-Hermitian framework. Also, the experimental realization of
related non-Hermitian systems is presented in refs. \cite{Franz,Ueda}.

In this paper, we introduce an unequal real amplitude of pairing operator to
describe the imbalance pair creation and annihilation in Kitaev chain. It is
a non-Hermitian model with time-reversal symmetry, rather than a combined $PT
$\ symmetry. Exact solution shows a rich phase diagram, including gapped
superconducting phases associated time-reversal symmetry and gapless phases
associated broken time-reversal symmetry. There are two topological
superconducting phases characterized by the extended Zak phase, which is
defined in the context of biorthonormal inner product. The boundary of these
two phases corresponds to the two-fold degeneracy, which is similar to the
Hermitian situation. In contrast, the gapless phase arises from coalescing
state in the broken symmetric regions, which indicates that the pairing
imbalance in some extent can destroy the superconducting phases. We also
investigate the bulk-edge correspondence in such a non-Hermitian system,
via\ the Majorana transformation. We show that there are two types of gapped
phases support different kinds of edge modes in the context of biorthonormal
inner product, which demonstrate the topological invariants.

This paper is organized as follows. In Section \ref{Non-Hermitian Kitaev
model}, we present the model Hamiltonian. Based on the solutions, we
investigate the phase diagram and analyze the symmetry of the ground state.
In Section \ref{Phase diagram and topological invariant}, we construct the
Hermitian counterpart of the model and calculate the Zak phase to show the
topological properties of the system. In Section \ref{Majorana edge modes},
we study the corresponding Hamiltonian of the Majorana fermions to show the
existence of the bulk-edge correspondence. Finally, we give a summary and
discussion in Section \ref{sec_summary}.

\begin{figure*}[tbp]
\includegraphics[ bb=78 50 546 512, width=0.45\textwidth, clip]{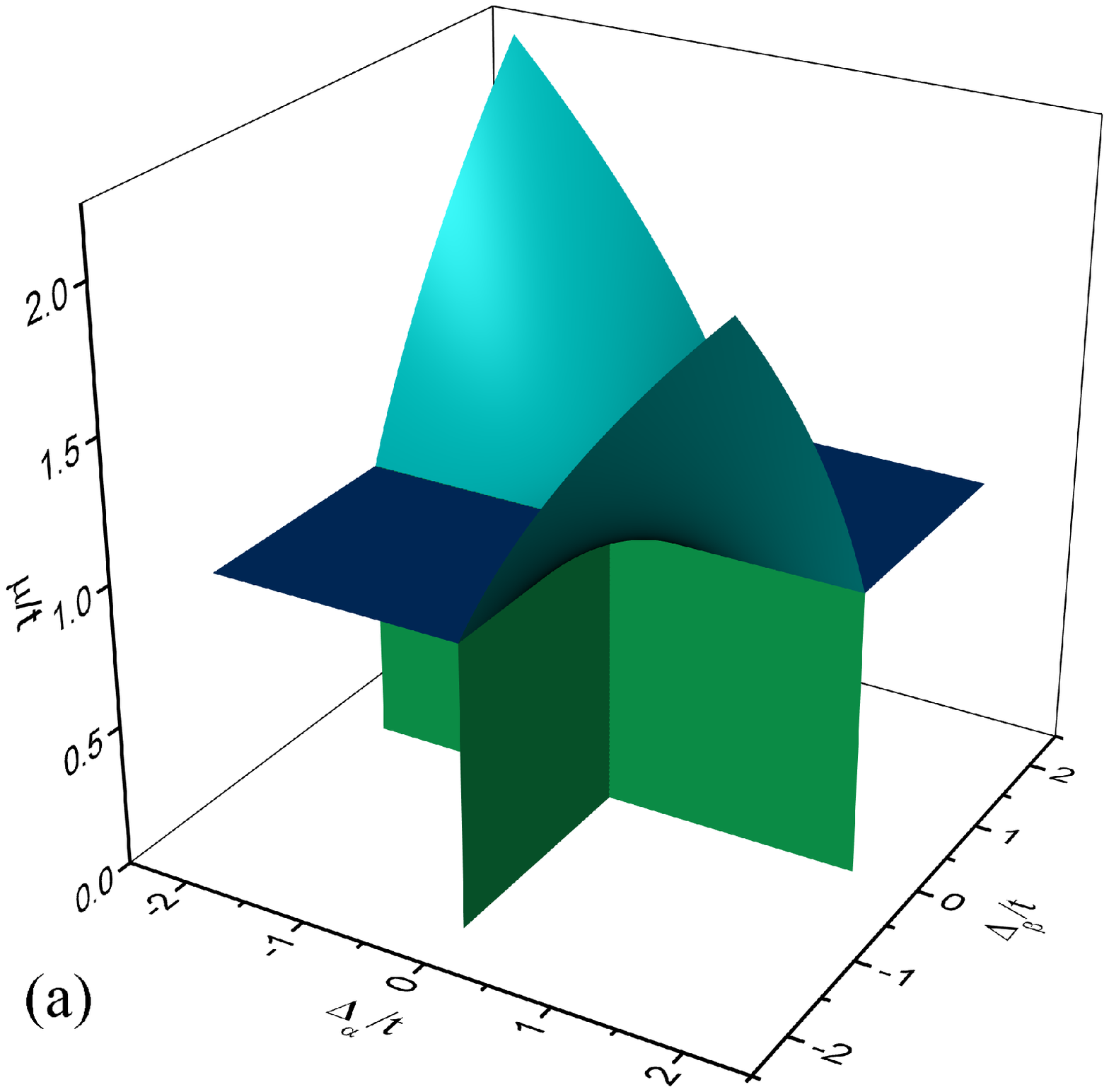} %
\includegraphics[ bb=78 50 546 512, width=0.45\textwidth, clip]{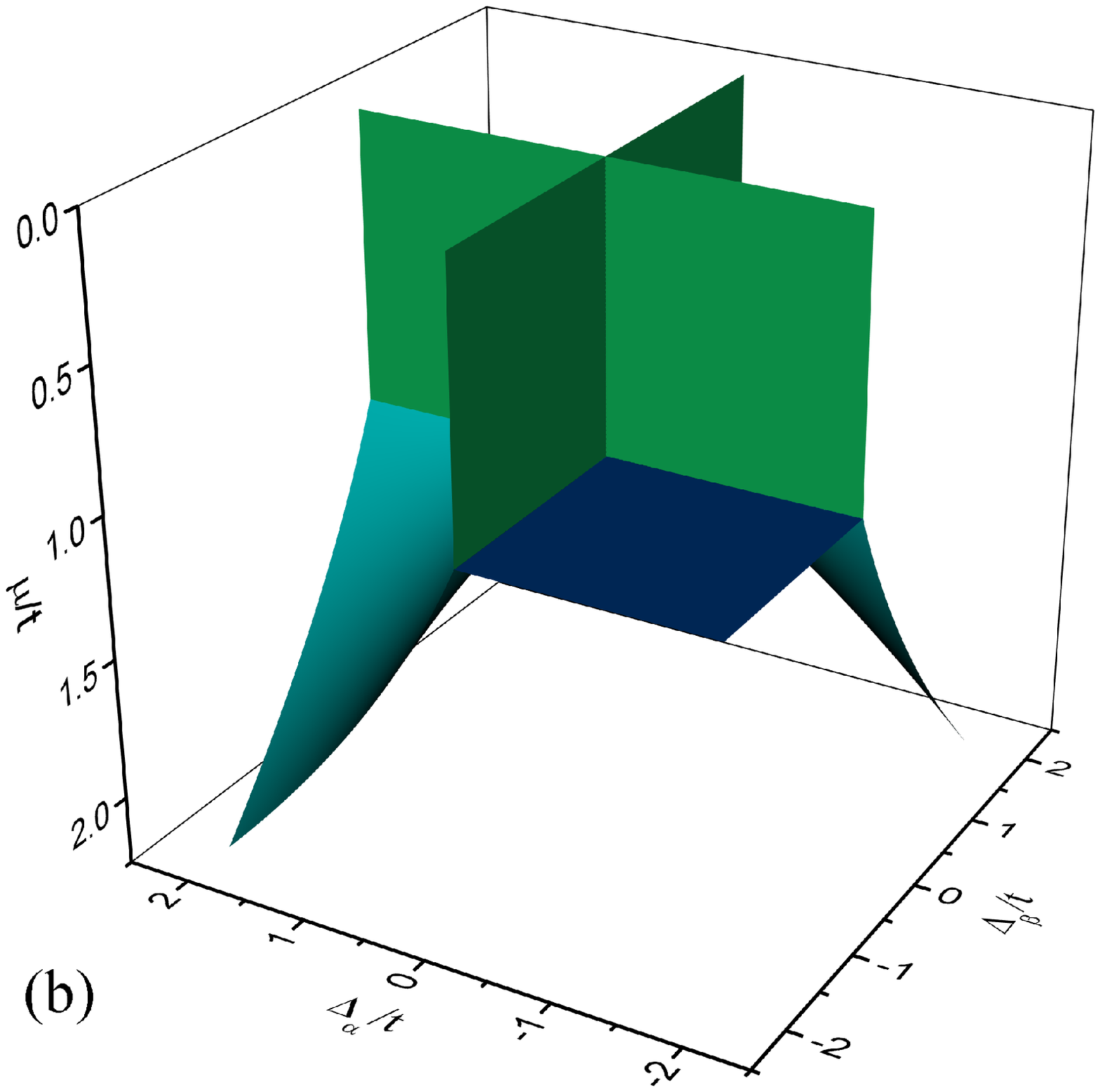}
\caption{(Color online) Schematic illustration of the diagram boundaries in
3D parameter space described in Table \protect\ref{Table I}. Three color
surfaces represent different interfaces. We only plot the diagram in
positive half-space since it is symmetric about $\protect\mu /t=0$\ plane.
Panels (a) and (b) are the same diagram seen from two different angles.}
\label{fig1}
\end{figure*}

\section{Non-Hermitian Kitaev model}

\label{Non-Hermitian Kitaev model} We consider the following fermionic
Hamiltonian on a lattice of length $N$

\begin{eqnarray}
\mathcal{H} &=&-t\sum\limits_{j=1}^{N}(c_{j}^{\dag }c_{j+1}+\text{\textrm{%
H.c.}})-\mu \sum\limits_{j=1}^{N}\left( 1-2n_{j}\right)  \notag \\
&&-\sum\limits_{j=1}^{N}(\Delta _{\alpha }c_{j}^{\dag }c_{j+1}^{\dag
}+\Delta _{\beta }c_{j+1}c_{j}),  \label{H}
\end{eqnarray}%
where $c_{j}^{\dag }$ $(c_{j})$\ is a fermionic creation (annihilation)
operator on site $j$, $n_{j}=c_{j}^{\dag }c_{j}$, $t$ the tunneling rate, $%
\mu
$ the chemical potential, and $\Delta _{\alpha }$ $(\Delta _{\beta })$\ the
strength of the $p$-wave pair creation and annihilation. For a closed chain,
we define $c_{N+1}=c_{1}$ and for an open chain, we set $c_{N+1}=0$. It
turns out that Hamiltonian (\ref{H}) has a rich phase diagram in its
Hermitian version, i.e., $\Delta _{\alpha }=\Delta _{\beta }$.

Before solving the Hamiltonian, it is profitable to investigate the symmetry
of the system and its breaking in the eigenstates. By the direct derivation,
we have $\left[ \mathcal{T},\mathcal{H}\right] =0$, i.e., the Hamiltonian is
a time reversal ($\mathcal{T}$) invariant, where the antilinear time
reversal operator $\mathcal{T}$ has the function $\mathcal{T}i\mathcal{T=-}i$%
. Before we consider the general non-Hermitian Kitaev model, we first
highlight the key idea for a limiting case. When $t=\mu =0$ and $\Delta
_{\alpha }=-\Delta _{\beta }$, the Hamiltonian (\ref{H}) reduces to

\begin{equation}
\mathcal{H}_{0}=\Delta _{\beta }\sum\limits_{j=1}^{N}(c_{j}^{\dag
}c_{j+1}^{\dag }-c_{j+1}c_{j}),
\end{equation}%
which is an anti-Hermitian Hamiltonian. Obviously, all the eigenvalues are
imaginary and break the $\mathcal{T}$\ symmetry, i.e., for an eigenstate $%
\left\vert \phi \right\rangle $\ with the nonzero eigenvalue $E$, $\mathcal{H%
}_{0}\left\vert \phi \right\rangle =E\left\vert \phi \right\rangle ,$ we have%
\begin{equation}
\mathcal{H}_{0}(\mathcal{T}\left\vert \phi \right\rangle )=-E(\mathcal{T}%
\left\vert \phi \right\rangle ).
\end{equation}%
It strongly implies that the $\mathcal{T}$\ symmetry in the present model
plays the same role as $\mathcal{PT}$\ symmetry\ in a $\mathcal{PT}$\
pseudo-Hermitian system. It motivates further study of such a model
systematically. Taking the Fourier transformation

\begin{equation}
c_{j}=\frac{1}{\sqrt{N}}\sum\limits_{k}e^{ikj}c_{k}
\end{equation}%
for the Hamiltonian, with wave vector $k=2\pi m/N$, $m=0,1,2,...,N-1$, we
have
\begin{eqnarray}
H &=&\sum\limits_{k}[2\left( \mu -t\cos k\right) c_{k}^{\dag }c_{k} \\
&&+i\sin k\left( \Delta _{\beta }c_{-k}c_{k}+\Delta _{\alpha }c_{-k}^{\dag
}c_{k}^{\dag }\right) -\mu ],  \notag
\end{eqnarray}%
So far the procedures are the same as those for solving the Hermitian
version of $H$. To diagonalize a non-Hermitian Hamiltonian, we should
introduce the Bogoliubov transformation in the complex version:%
\begin{eqnarray}
A_{k} &=&i\sqrt{\Delta _{\beta }/\Delta _{\alpha }}\xi c_{k}+\eta
c_{-k}^{\dag }, \\
\overline{A}_{k} &=&-i\sqrt{\Delta _{\alpha }/\Delta _{\beta }}\xi
c_{k}^{\dagger }+\eta c_{-k}  \notag
\end{eqnarray}%
where%
\begin{eqnarray}
\xi &=&\mathrm{sgn}(\sin k)\sqrt{\frac{\mu -t\cos k+\varepsilon _{k}}{%
2\varepsilon _{k}}},  \notag \\
\eta &=&\frac{\left\vert \sin k\right\vert \sqrt{\Delta _{\alpha }\Delta
_{\beta }}}{\sqrt{2\varepsilon _{k}\left( \varepsilon _{k}+\mu -t\cos
k\right) }}.
\end{eqnarray}

We would like to point out that this is the crucial step to solve the
non-Hermitian Hamiltonian, which essentially establish the biorthogonal
bases. Obviously, complex Bogoliubov modes $\left( A_{k},\overline{A}%
_{k}\right) $ satisfy the canonical commutation relations%
\begin{eqnarray}
\left\{ A_{k},\overline{A}_{k^{\prime }}\right\} &=&\delta _{k,k^{\prime }},
\\
\left\{ A_{k},A_{k^{\prime }}\right\} &=&\left\{ \overline{A}_{k},\overline{A%
}_{k^{\prime }}\right\} =0,  \notag
\end{eqnarray}%
which result in the diagonal form of the Hamiltonian%
\begin{equation}
H=\sum\limits_{k}\epsilon _{k}\left( \overline{A}_{k}A_{k}-\frac{1}{2}%
\right) .
\end{equation}%
Here the single-particle spectrum in each subspace is%
\begin{equation}
\epsilon _{k}=2\sqrt{\left( \mu -t\cos k\right) ^{2}+\Delta _{\alpha }\Delta
_{\beta }\sin ^{2}k}.
\end{equation}%
Note that the Hamiltonian $H$\ is still non-Hermitian due to the fact that $%
\overline{A}_{k}\neq A_{k}^{\dag }$. Accordingly, the eigenstates of $H$ can
be written as the form%
\begin{equation}
\prod\limits_{\left\{ k\right\} }\overline{A}_{k}\left\vert \text{G}%
\right\rangle ,
\end{equation}%
which constructs the biorthogonal set associated with the eigenstates%
\begin{equation}
\left\langle \overline{\text{G}}\right\vert \prod\limits_{\left\{ k\right\}
}A_{k}
\end{equation}%
of the Hamiltonian $H^{\dag }$, where $\left\vert \text{G}\right\rangle \
(\left\langle \overline{\text{G}}\right\vert )$ is the ground state of $H$ ($%
H^{\dag }$). In the following, we will investigate the phase diagram based
on the properties of the solutions.

It is clear that when any one of the momentum $k$\ satisfies%
\begin{equation}
\left( \mu -t\cos k\right) ^{2}+\Delta _{\alpha }\Delta _{b}\sin ^{2}k<0,
\end{equation}%
the imaginary energy level appears in single-particle spectrum, which leads
to the occurrence of complex energy level for the Hamiltonian (\ref{H}), and
the $\mathcal{T}$ symmetry is broken in the corresponding eigenstates. This
can be seen from the properties of the single-particle spectrum and the
ground states $\left\vert \text{G}\right\rangle $.

Firstly, we focus on the boundary between the broken and unbroken symmetry
regions, as well as between two gapped phases. In the thermodynamical limit,
the boundary is a 2D surface in 3D parameter space $\left( \Delta _{\alpha
}/t,\Delta _{\beta }/t,\mu /t\right) $, which is determined by equation $%
\epsilon _{k}=0$. A straightforward algebra gives the analytical expression
of surfaces as the phase boundaries, which are listed in Table \ref{Table I}%
, and is plotted in Fig. \ref{fig1} as illustration. Note that there exist
two kinds of boundaries which consist of exceptional point (EP) and
degeneracy point, respectively.

\begin{table}[tbp]
\caption{Quantum phase boundaries in parameter space and corresponding
topological invariants, which is also illustrated in Fig. \protect\ref{fig1}
and \protect\ref{fig2}.} \label{Table I}\renewcommand\arraystretch{1}

\begin{tabular}{ccc}
\hline\hline
Regions & Phase & Zak phase \\ \hline
$\Delta _{\alpha }/t,\Delta _{\beta }/t>0,\left\vert \mu /t\right\vert <1$ &
Gapped & $-\pi $ \\
$\Delta _{\alpha }/t,\Delta _{\beta }/t<0,\left\vert \mu /t\right\vert <1$ &
Gapped & $\pi $ \\
$\Delta _{\alpha }\Delta _{\beta }/t^{2}>0,\mu /t=\pm 1$ & Gapless &  \\
$\left\vert \mu /t\right\vert >1,\mu ^{2}+\Delta _{\alpha }\Delta _{\beta
}>t^{2}$ & Gapped & $0$ \\
Otherwise & Coalescing &  \\ \hline\hline
\end{tabular}
\end{table}

Secondly, according to the non-Hermitian quantum theory, the occurrence of
the EP always accomplishes the $\mathcal{T}$\ symmetry breaking of an
eigenstate. For the present model, the symmetry of the groundstate $%
\left\vert \text{G}\right\rangle $ can be an indicator of the phase
transition due to the fact that the groundstate energy becomes complex once
the system is in the broken region. In the following, we focus on the
discussion about the symmetry of $\left\vert \text{G}\right\rangle $\ in the
different regions.

Applying the $\mathcal{T}$\ operator on the fermion operators and its vacuum
state $\left\vert \text{Vac}\right\rangle $, we have%
\begin{equation}
\mathcal{T}c_{k}^{\dagger }\left( \mathcal{T}\right) ^{-1}=c_{-k}^{\dagger },%
\mathcal{T}c_{k}\left( \mathcal{T}\right) ^{-1}=c_{-k},
\end{equation}%
and

\begin{equation}
\mathcal{T}\left\vert \text{Vac}\right\rangle =\left\vert \text{Vac}%
\right\rangle ,
\end{equation}%
which are available in the both regions. However, the coefficients $\xi $
and $\eta $\ experience a transition as following when the corresponding
single-particle level changes from real to imaginary: We have $\xi ^{\ast }$
$=\xi $ and $\eta ^{\ast }$ $=\eta $ for real levels and $\xi ^{\ast }$ $=%
\mathrm{sgn}(\sin k)\eta $ for the imaginary levels, respectively. This
leads to the conclusion that the groundstate is not $\mathcal{T}$ symmetric
in the broken symmetric region, i.e.,%
\begin{equation}
\left\{
\begin{array}{cc}
\mathcal{T}\left\vert \text{G}\right\rangle =\left\vert \text{G}%
\right\rangle ; & \text{Gapped region} \\
\mathcal{T}\left\vert \text{G}\right\rangle \neq \left\vert \text{G}%
\right\rangle , & \text{Gapless region}%
\end{array}%
\right. .
\end{equation}%
It shows that the $\mathcal{T}$\ symmetry in the present model plays the
same role as $\mathcal{PT}$\ symmetry\ in a $\mathcal{PT}$\ pseudo-Hermitian
system.

As a comparison, it is noted that the phase boundary $\Delta _{\alpha
}=\Delta _{\beta }=0$ between two gapped phases is not an EP line but the
degeneracy line. So across the $\mu $\ axis, the quantum phases transition
is a conventional one as that in a Hermitian Kitaev chain. We will
demonstrate this point in the next section.

\begin{figure*}[tbp]
\includegraphics[ bb=36 248 373 537, width=0.3\textwidth, clip]{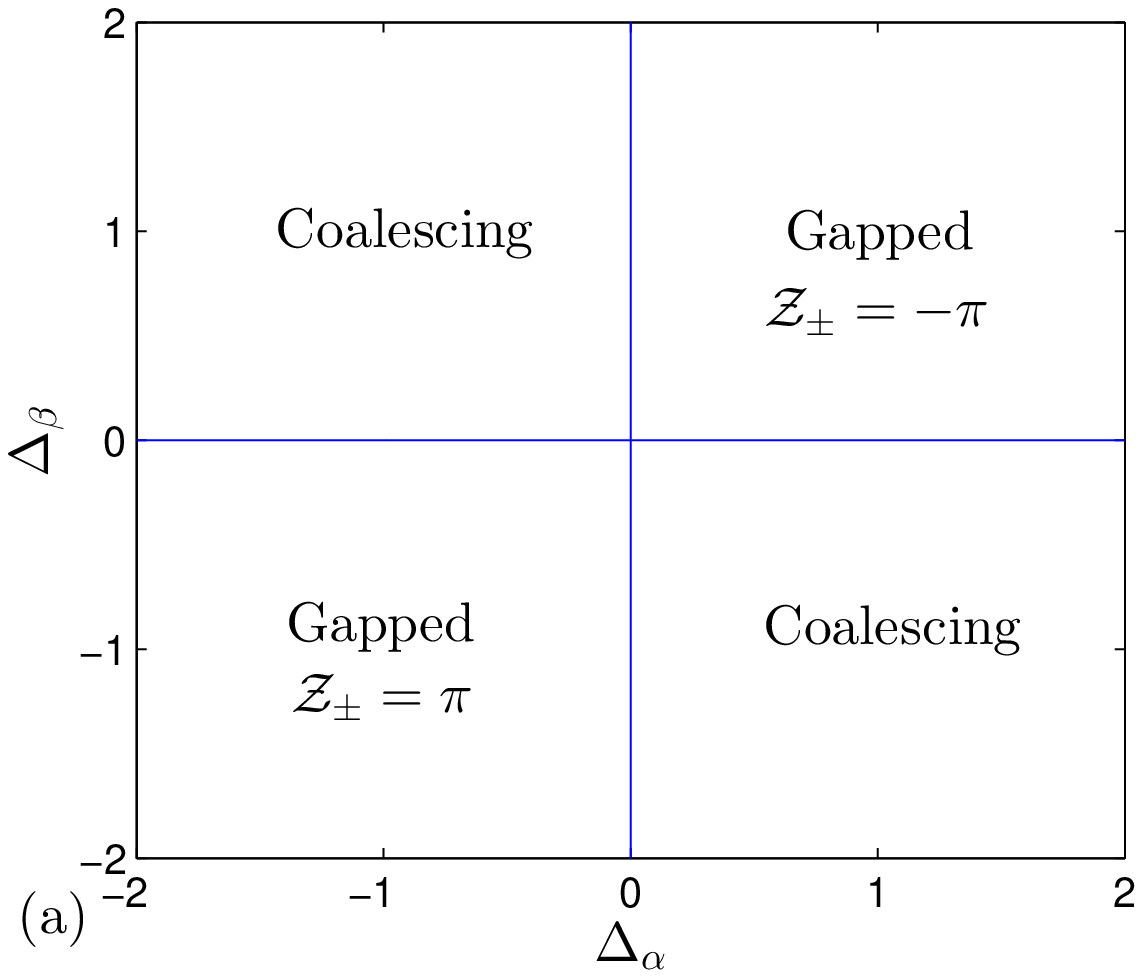} %
\includegraphics[ bb=36 248 373 537, width=0.3\textwidth, clip]{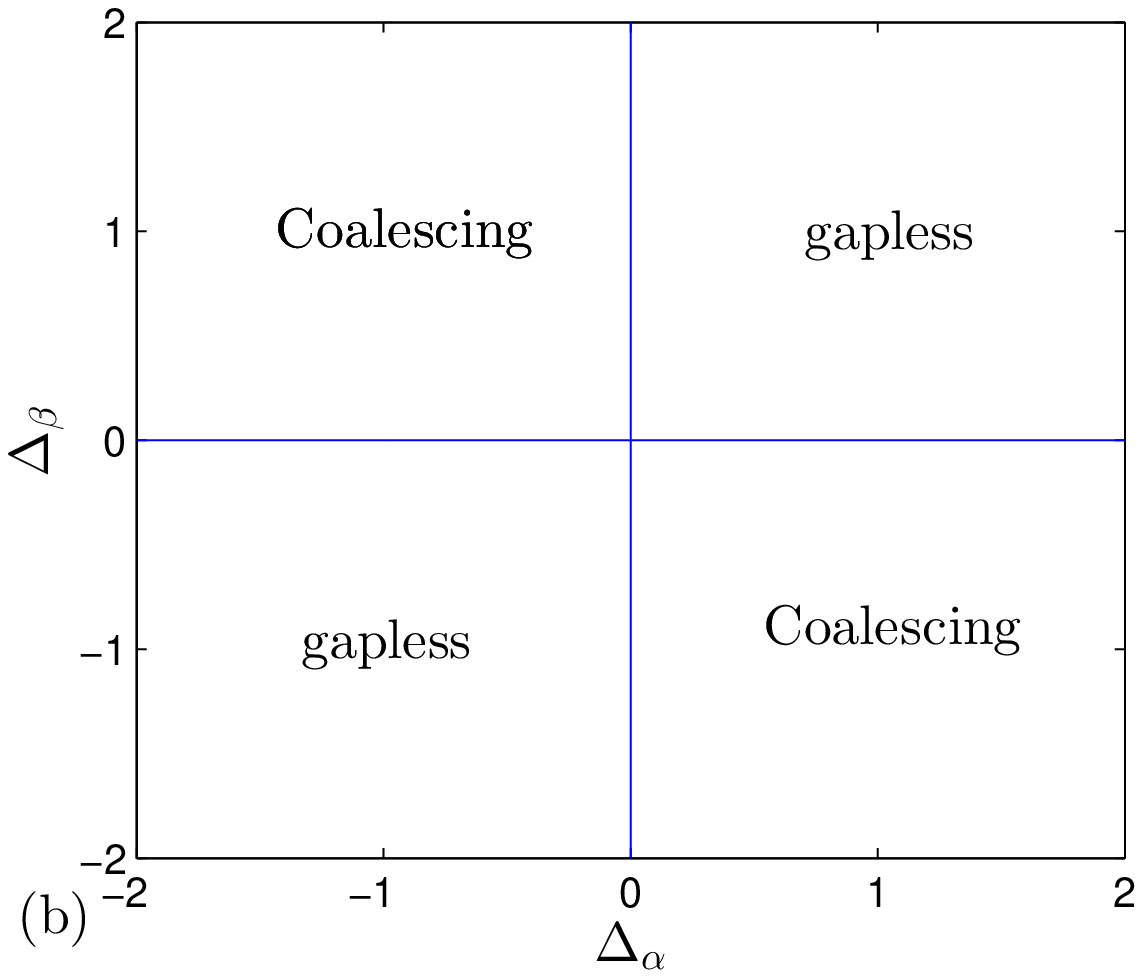} %
\includegraphics[ bb=36 248 373 537, width=0.3\textwidth, clip]{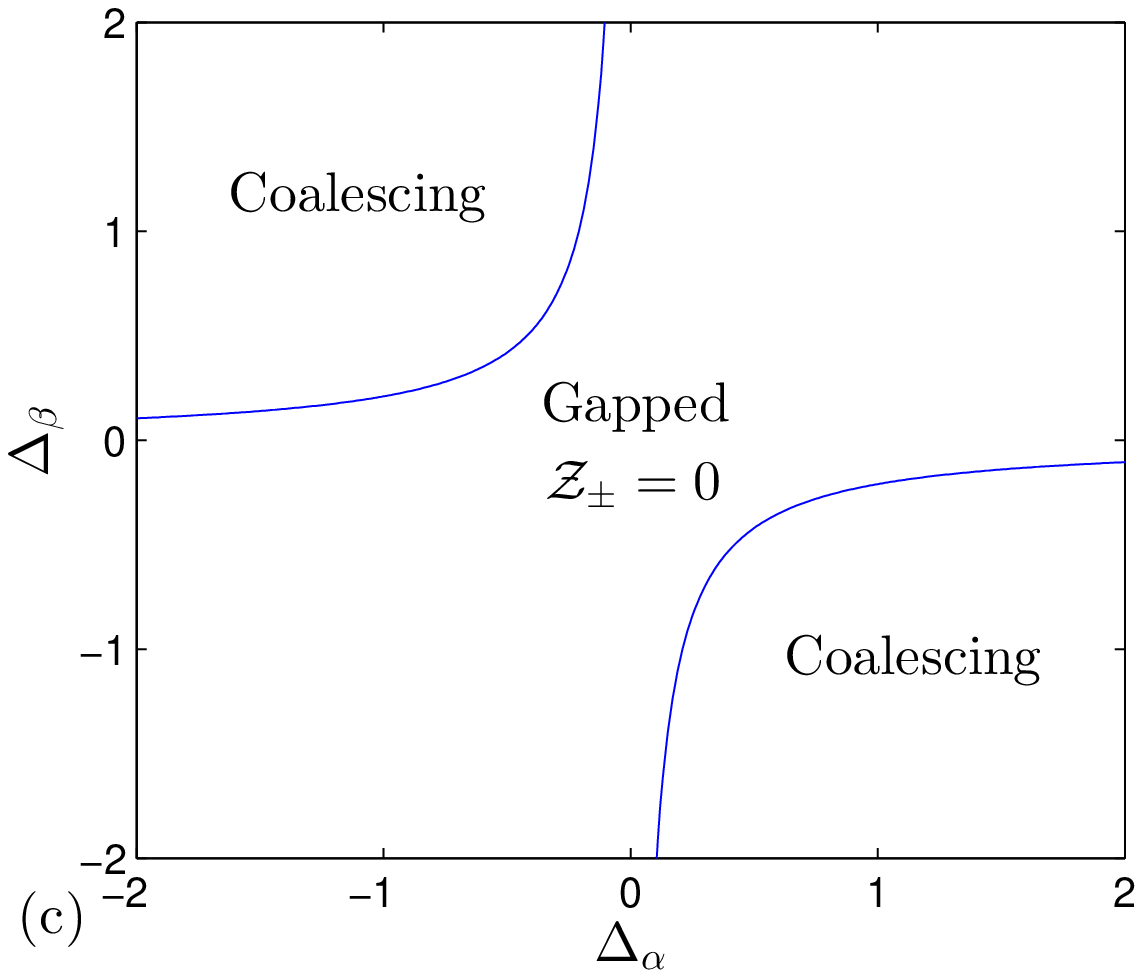}
\caption{(Color online) Phase diagrams of the imbalanced kitaev model in $%
\Delta _{\protect\alpha }\Delta _{\protect\beta }$ plane for several
representative values of (a) $\protect\mu /t=\pm 0.5$, (b) $\protect\mu %
/t=\pm 1$ and (c) $\protect\mu /t=\pm 1.1$. The phase descriptions and
corresponding Zak phases are indicated in the regions.}
\label{fig2}
\end{figure*}

\section{Phase diagram and topological invariant}

\label{Phase diagram and topological invariant} According to the above
analysis, the whole parameter space consists of two kinds of regions,
unbroken symmetric one with fully real spectrum and broken symmetric one
with complex spectrum, which is originated from the imbalanced pairing
process. In this section, we focus on the former region and show that it is
further divided into three kinds of sub-regions\ with different topological
invariants. Using the Nambu representation, the Hamiltonian $\mathcal{H}$
can be expressed as

\begin{equation}
\mathcal{H}=\sum\limits_{k}(c_{k}^{\dag },c_{-k})h_{k}\left(
\begin{array}{c}
c_{k} \\
c_{-k}^{\dag }%
\end{array}%
\right)
\end{equation}%
where the core matrix is%
\begin{equation}
h_{k}=\left(
\begin{array}{cc}
\mu -t\cos k & -i\Delta _{\alpha }\sin k \\
i\Delta _{\beta }\sin k & -\mu +t\cos k%
\end{array}%
\right) .
\end{equation}%
Matrix $h_{k}$\ contains all the information of the system. (i) In the
unbroken symmetric region, $h_{k}$\ is a pseudo-Hermitian matrix, possessing
a Hermitian counterpart%
\begin{equation}
\widetilde{h}_{k}=\left(
\begin{array}{cc}
\mu -t\cos k & -i\sqrt{\Delta _{\alpha }\Delta _{\beta }}\sin k \\
i\sqrt{\Delta _{\alpha }\Delta _{\beta }}\sin k & -\mu +t\cos k%
\end{array}%
\right) ,
\end{equation}%
which is a Hermitian matrix with the same real eigenvalues of $h_{k}$. (ii)
In the broken symmetric region, there are several $k_{c}$ satisfing\
\begin{equation}
\cos k_{c}=\frac{\mu t\pm \sqrt{\Delta _{\alpha }\Delta _{\beta }\left( \mu
^{2}+\Delta _{\alpha }\Delta _{\beta }-t^{2}\right) }}{t^{2}-\Delta _{\alpha
}\Delta _{\beta }},
\end{equation}%
\ at which $h_{k_{c}}$\ is equivalent to a Jordan block. Explicitly we have
\begin{equation}
sh_{k_{c}}s^{-1}=\left(
\begin{array}{cc}
0 & 1 \\
0 & 0%
\end{array}%
\right) ,
\end{equation}%
where
\begin{equation}
s=\left(
\begin{array}{cc}
1 & 0 \\
\mu -t\cos k_{c} & -i\Delta _{\alpha }\sin k%
\end{array}%
\right) .
\end{equation}%
Two eigenvectors of the Jordan block coalesce, resulting coalescing ground
state. This is an exclusive feature of the non-Hermitian system. (iii) Along
the $\mu $\ axis, there are two nodal points at $k_{c}=\pm \arccos \mu $,
within the region $\left\vert \mu \right\vert \leqslant 1$, at which the
matrix reduces to%
\begin{equation}
h_{k_{c}}=\left(
\begin{array}{cc}
0 & 0 \\
0 & 0%
\end{array}%
\right) .
\end{equation}%
Then there are two zero-energy eigen vectors which accords with the results
obtained directly from the Hermitian Hamiltonian with $\Delta _{\alpha
}=\Delta _{\beta }=0$. We will see that it is the line (rather than a
surface) that separates two topological phases.

In general, $h_{k}$\ can be expressed as $h_{k}=\mathbf{d(}k\mathbf{)}\cdot
\mathbf{\sigma }$, with a $3$D complex vector field%
\begin{equation}
\left\{
\begin{array}{l}
d_{x}\mathbf{(}k\mathbf{)}\mathbf{=-}\frac{i}{2}\left( \Delta _{\alpha
}-\Delta _{\beta }\right) \sin k, \\
d_{y}\mathbf{(}k\mathbf{)}\mathbf{=}\frac{1}{2}\left( \Delta _{\alpha
}+\Delta _{\beta }\right) \sin k, \\
d_{z}\mathbf{(}k\mathbf{)}\mathbf{=}\mu -t\cos k,%
\end{array}%
\right.
\end{equation}%
where $\mathbf{\sigma }$\ represents $3$D\ Pauli matrices.

Now we focus on the unbroken symmetry region ($\Delta _{\alpha }\Delta
_{\beta }>0$). The eigenstates of a pseudo-Hermitian Hamiltonian can
construct a complete set of biorthogonal bases in association with the
eigenstates of its Hermitian conjugate. For present system, eigenstates $%
\left\vert \psi _{+}^{k}\right\rangle $, $\left\vert \psi
_{-}^{k}\right\rangle $ of $h_{k}$ and $\left\vert \eta
_{+}^{k}\right\rangle $, $\left\vert \eta _{-}^{k}\right\rangle $ of $%
h_{k}^{\dagger }$ are the biorthogonal bases of the single-particle
invariant subspace, which are explicitly expressed as

\begin{equation}
\begin{array}{cc}
\left\vert \psi _{+}^{k}\right\rangle =\left(
\begin{array}{c}
\cos \frac{\theta }{2}e^{-i\phi } \\
\sin \frac{\theta }{2}%
\end{array}%
\right) , & \left\vert \psi _{-}^{k}\right\rangle =\left(
\begin{array}{c}
\sin \frac{\theta }{2}e^{-i\phi } \\
-\cos \frac{\theta }{2}%
\end{array}%
\right) , \\
&  \\
\left\vert \eta _{+}^{k}\right\rangle =\left(
\begin{array}{c}
\cos \frac{\theta }{2}e^{i\phi } \\
\sin \frac{\theta }{2}%
\end{array}%
\right) ^{\ast }, & \left\vert \eta _{-}^{k}\right\rangle =\left(
\begin{array}{c}
\sin \frac{\theta }{2}e^{i\phi } \\
-\cos \frac{\theta }{2}%
\end{array}%
\right) ^{\ast }.%
\end{array}%
\end{equation}%
By rewriting the field $\mathbf{d(}k\mathbf{)}$ as%
\begin{equation}
\mathbf{d(}k\mathbf{)}=r\left( \cos \theta ,\sin \theta \sin \phi ,\sin
\theta \cos \phi \right)
\end{equation}%
where%
\begin{eqnarray}
r &=&\sqrt{\left( \mu -t\cos k\right) ^{2}+\Delta _{\alpha }\Delta _{\beta
}\sin ^{2}k},  \notag \\
\cos \theta &=&\mathbf{-}\frac{i}{2r}\left( \Delta _{\alpha }-\Delta _{\beta
}\right) \sin k, \\
\tan \phi &=&\frac{\left( \Delta _{\alpha }+\Delta _{\beta }\right) \sin k}{%
2\left( \mu -t\cos k\right) }.  \notag
\end{eqnarray}%
It is ready to check that biorthogonal bases $\left\{ \left\vert \psi
_{\lambda }^{k}\right\rangle ,\left\vert \eta _{\lambda }^{k}\right\rangle
\right\} $ ($\lambda =\pm $) obey the biorthogonal and completeness
conditions
\begin{equation}
\langle \eta _{\lambda ^{\prime }}^{k^{\prime }}\left\vert \psi _{\lambda
}^{k}\right\rangle =\delta _{\lambda \lambda ^{\prime }}\delta _{kk^{\prime
}},\text{ }\sum_{\lambda ,k}\left\vert \psi _{\lambda }^{k}\right\rangle
\left\langle \eta _{\lambda }^{k}\right\vert =1.
\end{equation}%
These properties are independent of the reality of the spectrum except for
the EPs. In unbroken symmetric region, $h_{k}$\ has the fully real spectrum.
To characterize the property of the energy band, we introduce the extended
Zak phase%
\begin{equation}
\mathcal{Z}_{\pm }=\int_{0}^{2\pi }\mathcal{A}_{k}\mathrm{d}k,
\end{equation}%
where the Berry connection%
\begin{equation}
\mathcal{A}_{k}=\left\langle \eta _{\pm }^{k}\right\vert \frac{\partial }{%
\partial k}\left\vert \psi _{\pm }^{k}\right\rangle =\partial _{k}\phi
\mathcal{A}_{\phi }+\partial _{k}\theta \mathcal{A}_{\theta },
\end{equation}%
with%
\begin{equation}
\mathcal{A}_{\phi }=\left\langle \eta _{\pm }^{k}\right\vert i\partial
_{\phi }\left\vert \psi _{\pm }^{k}\right\rangle ,\mathcal{A}_{\theta
}=\left\langle \eta _{\pm }^{k}\right\vert i\partial _{\theta }\left\vert
\psi _{\pm }^{k}\right\rangle .
\end{equation}%
It is easy to check that%
\begin{equation}
\mathcal{Z}_{\pm }=\left\{
\begin{array}{cc}
-\pi \mathrm{sgn}\left( \frac{\Delta _{\alpha }+\Delta _{\beta }}{t}\right) ,
& \left\vert \mu \right\vert <1 \\
0, & \left\vert \mu \right\vert >1%
\end{array}%
\right. .
\end{equation}%
In Fig. \ref{fig2}, we show the phase diagram described in Table \ref{Table
I} with several typical values of $\mu /t$, in which the corresponding Zak
phases are indicated as topological invariants.

\begin{figure}[tbp]
\includegraphics[ bb=22 234 373 763, width=0.45\textwidth, clip]{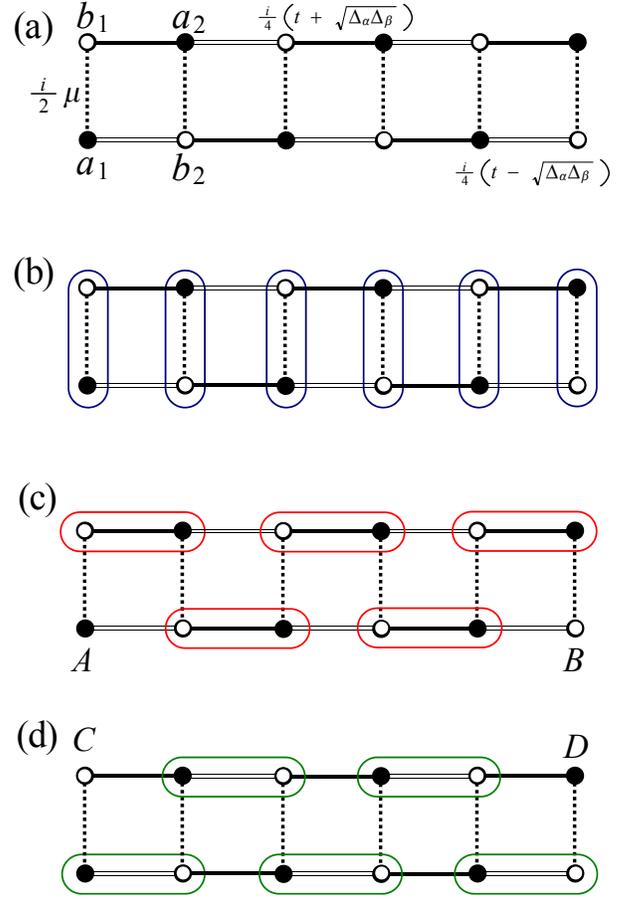} %
\caption{(Color online) Schematic structure and the formation of edge modes
of the Majorana Hamiltonian from Eq. (\protect\ref{majorana}), which can be
regarded as two coupled SSH chains. The system consists of two sublattices $a $ and $b$, indicated by filled and empty circles, respectively (a) Hopping
amplitudes along each chain are staggered by $i(t-\protect\sqrt{\Delta _{\protect\alpha }\Delta _{\protect\beta }})/4$ (double line) and $i(t+\protect\sqrt{\Delta _{\protect\alpha }\Delta _{\protect\beta }})/4$ (single line).
The interchain hopping amplitude is $i\protect\mu /2$ (dotted line). (b) In
the case of $\left\vert \protect\mu \right\vert \gg \left\vert t-\protect\sqrt{\Delta _{\protect\alpha }\Delta _{\protect\beta }}\right\vert$, the
interchain dimerization (circled by the blue line) results in topologically
trivial phase, since there is no zero energy edge state. (c) In the limit $\left\vert t+\protect\sqrt{\Delta _{\protect\alpha }\Delta _{\protect\beta }}\right\vert \gg \left\vert t-\protect\sqrt{\Delta _{\protect\alpha }\Delta _{\protect\beta }}\right\vert$, edge states form at sites $A$ and $B$ due to
the dimerization (circled by the red line). (d) Edge states form at sites $C$
and $D$ in the case $\left\vert t-\protect\sqrt{\Delta _{\protect\alpha }\Delta _{\protect\beta }}\right\vert \gg \left\vert t+\protect\sqrt{\Delta
_{\protect\alpha }\Delta _{\protect\beta }}\right\vert$, due to the
dimerization (circled by the green line). These results accord with the bulk
phase diagram in Fig. \protect\ref{fig2} for unbroken symmetric regions.} %
\label{fig3}
\end{figure}

\begin{figure*}[tbp]
\includegraphics[ bb=25 242 362 530, width=0.45\textwidth, clip]{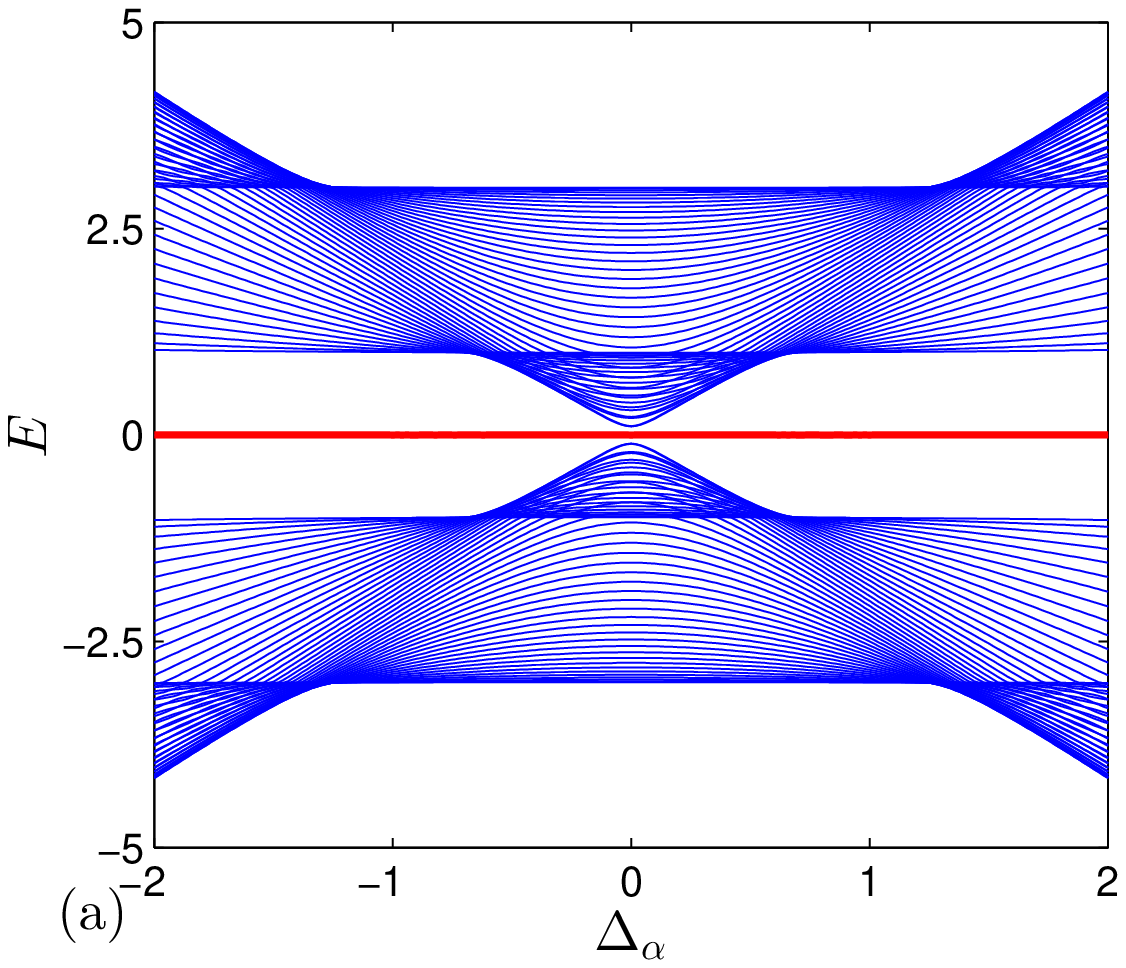} %
\includegraphics[ bb=25 242 362 530, width=0.45\textwidth, clip]{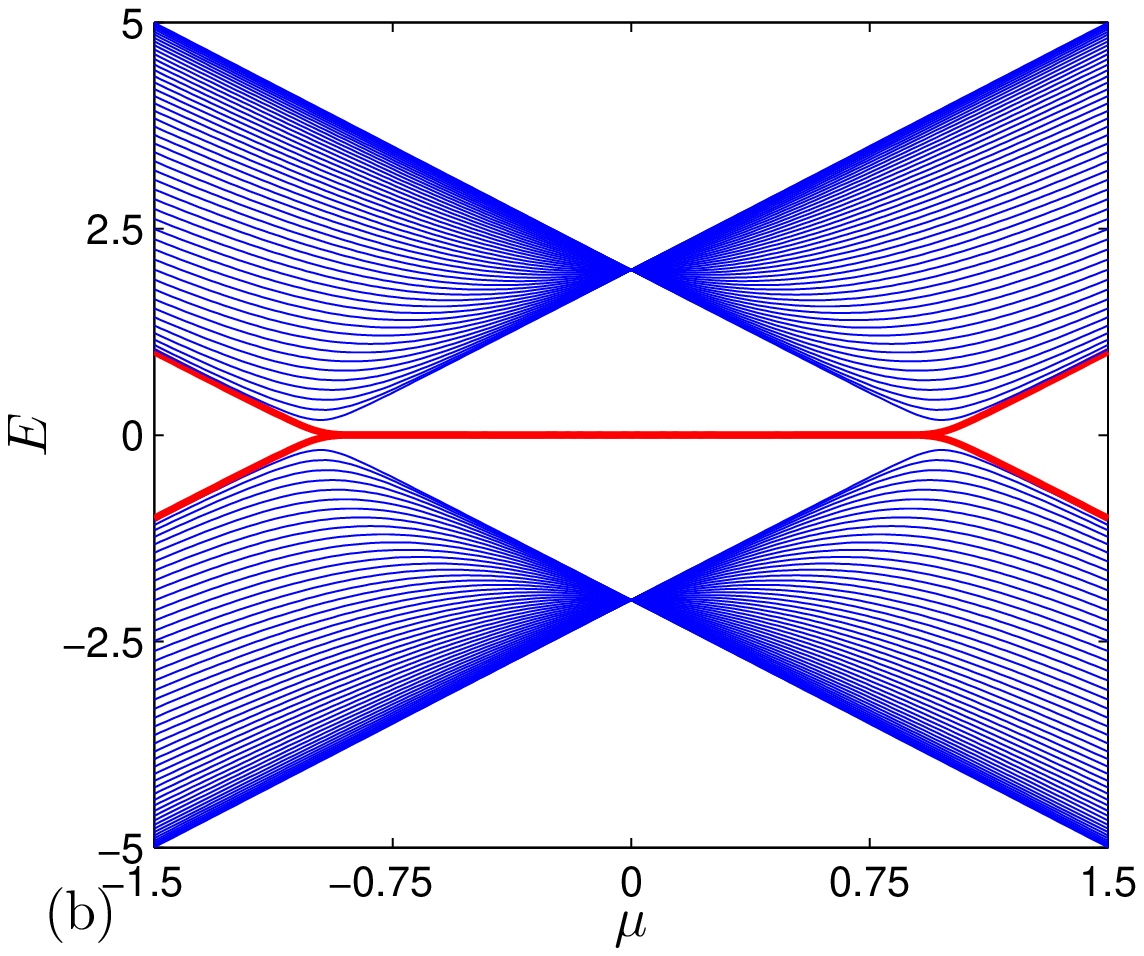}
\caption{(Color online) Energy spectra for the Hamiltonian in Eq. (\protect
\ref{majorana}) as functions of $\protect\sqrt{\Delta _{\protect\alpha %
}\Delta _{\protect\beta }}$ and $\protect\mu $ in the unbroken symmetric
regions on a $2\times 50$ ladder with the open boundary condition, obtained
by exact diagonalization. The parameters are (a) $\protect\mu =0.5$, $\Delta
_{\protect\alpha }=\Delta _{\protect\beta }$ and (b) $\Delta _{\protect%
\alpha }=\Delta _{\protect\beta }=-1$. Here the energy is in the unit of $%
1/4 $ and $t=1$. The red lines indicates the midgap levels. We see that
energy gaps open when $\protect\mu >1$, while the zero modes appear in the
region of $\protect\mu <1$, which accords with the Zak phases in both
regions.}
\label{fig4}
\end{figure*}

\section{Bulk-edge correspondence}

\label{Majorana edge modes}

Based on the above analysis, it turns out that the bulk system exhibits the
similar topological feature within the unbroken symmetric regions. In
Hermitian systems, the existence of edge modes is intimately related to the
bulk topological quantum numbers, which is referred as the bulk-edge
correspondence relations \cite{Thouless,Kane,Zhang,Lu}. We are interested in
the generalization of the bulk-edge correspondence to non-Hermitian systems.
Previous works show that when sufficiently weak non-Hermiticity is
introduced to topological insulator models, the edge modes can retain some
of their original characteristics \cite{Esaki,Hu}.

We examine the bulk-edge correspondence in the present model based on the
following strategy. It is well known that there is a Hermitian counterpart
for a non-Hermitian Hamiltonian within the unbroken symmetric region, in
which both two Hamiltonians have the identical fully real spectrum. It
allows us to find out the zero modes of the non-Hermitian Hamiltonian with
the open boundary condition from its Hermitian counterpart through Majorana
transformation.

Before we focus on the concrete model, we first present a general formalism
for a non-Hermitian Kitaev model within the unbroken symmetric region.%
\textbf{\ }We write down a generic Hamiltonian in the Bogoliubov-de Gennes
formalism%
\begin{equation}
\mathcal{H}=\frac{1}{2}\overline{D}h_{\mathrm{BdG}}D,
\end{equation}%
with $D$\ and $\overline{D}$\ column vectors containing all biorthonormal
canonical operators,\textbf{\ }$D=(d_{1},\ldots ,d_{N},$ $\overline{d}%
_{1},\ldots ,\overline{d}_{N})^{T}$\textbf{, }$\overline{D}=(\overline{d}%
_{1},\ldots ,\overline{d}_{N},$ $d_{1},\ldots ,d_{N})$\textbf{.} Here, the
canonical operators are defined as \cite{ZXZ1}%
\begin{equation}
d_{j}=\sqrt{\frac{\Delta _{\beta }}{\Delta _{\alpha }}}c_{j},\overline{d}%
_{j}=\sqrt{\frac{\Delta _{\alpha }}{\Delta _{\beta }}}c_{j}^{\dagger },
\label{cd}
\end{equation}%
which satisfy the canonical relations%
\begin{equation}
\{d_{j},\overline{d}_{l}\}=\delta _{jl},\{d_{j},d_{l}\}=\{\overline{d}_{j},%
\overline{d}_{l}\}=0.  \notag
\end{equation}%
Here $h_{\mathrm{BdG}}$\ is a Hermitian matrix and then one can construct a
Hamiltonian%
\begin{equation}
H_{\mathrm{cp}}\mathcal{=}\frac{1}{2}C^{\dag }h_{\mathrm{BdG}}C,
\end{equation}%
which represents a Hermitian version of the Kitaev Hamiltonian.\textbf{\ }$%
H_{\mathrm{cp}}$\textbf{\ }possesses the identical spectrum with that of $%
\mathcal{H}$, and is dubbed as the Hermitian counterpart of $\mathcal{H}$
\cite{Ali}.\ This ensures that one can map some results of $H_{\mathrm{cp}}$%
\ to $\mathcal{H}$\ directly. For instance, we have the following relation:
For an arbitrary compound operator $g(C^{\dag },C)$, which satisfies%
\begin{equation}
\lbrack g(C^{\dag },C),H_{\mathrm{cp}}]=0,  \label{eq1}
\end{equation}%
we always have%
\begin{equation}
\lbrack g(\overline{D},D),\mathcal{H}]=0.  \label{eq2}
\end{equation}%
We refer the mapping $[g(C^{\dag },C),H_{\mathrm{cp}}]\rightarrow \lbrack g(%
\overline{D},D),\mathcal{H}]$\ as an equivalence relation between $\mathcal{H%
}$\ and $H_{\mathrm{cp}}$, which will be used in the subsequent zero-mode
analysis.

Now we apply the formalism to the non-Hermitian Kitaev Hamiltonian. In the
rest, we only focus on the unbroken regions. Using the transformation in Eq.
(\ref{cd}) we rewrite the Hamiltonian with the open boundary condition as%
\begin{eqnarray}
\mathcal{H} &=&-t\sum\limits_{j=1}^{N-1}(\overline{d}_{j}d_{j+1}+\overline{d}%
_{j+1}d_{j})-\mu \sum\limits_{j=1}^{N}\left( 1-2\overline{n}_{j}\right)
\notag \\
&&-\sqrt{\Delta _{\alpha }\Delta _{\beta }}\sum\limits_{j=1}^{N-1}(\overline{%
d}_{j}\overline{d}_{j+1}+d_{j+1}d_{j}),  \label{HD}
\end{eqnarray}%
where\textbf{\ }$\overline{n}_{j}=\overline{d}_{j}d_{j}$ is the particle
number operator in the context of biorthonormal inner product.\textbf{\ }The
Hermitian counterpart of $\mathcal{H}$\ can be obtained by replacing $(%
\overline{d}_{j},d_{j})$\ with $\left( c_{j}^{\dagger },c_{j}\right) $, i.e.,%
\begin{eqnarray}
H_{\mathrm{cp}} &=&-t\sum\limits_{j=1}^{N-1}(c_{j}^{\dag }c_{j+1}+\text{%
\textrm{H.c.}})-\mu \sum\limits_{j=1}^{N}\left( 1-2n_{j}\right)  \notag \\
&&-\sqrt{\Delta _{\alpha }\Delta _{\beta }}\sum\limits_{j=1}^{N-1}(c_{j}^{%
\dag }c_{j+1}^{\dag }+\text{\textrm{H.c.}})  \label{HC}
\end{eqnarray}%
with\textbf{\ }$n_{j}=c_{j}^{\dagger }c_{j}$. Next we concentrate on the
Hamiltonian $H_{\mathrm{cp}}$, and then apply the obtained result to $%
\mathcal{H}$. We induce the Majorana operators%
\begin{equation}
a_{j}=c_{j}^{\dagger }+c_{j},b_{j}=-i\left( c_{j}^{\dagger }-c_{j}\right) ,
\end{equation}%
which satisfy the relations%
\begin{eqnarray}
\{a_{j},a_{l}\} &=&2\delta _{jl},\{b_{j},b_{l}\}=2\delta _{jl},  \notag \\
\{a_{j},b_{l}\} &=&0,a_{j}^{2}=b_{j}^{2}=1.
\end{eqnarray}%
Then the Majorana representation of $H_{\mathrm{cp}}$\ is%
\begin{eqnarray}
&&H_{\mathrm{cp}}=-\frac{1}{4}\{\sum\limits_{j=1}^{N-1}[i\left( t+\sqrt{%
\Delta _{\alpha }\Delta _{\beta }}\right) b_{j}a_{j+1}  \notag \\
&&+i\left( t-\sqrt{\Delta _{\alpha }\Delta _{\beta }}\right)
b_{j+1}a_{j}]+i2\mu \sum\limits_{j=1}^{N}a_{j}b_{j}+\text{\textrm{H.c.}}\}.
\label{majorana}
\end{eqnarray}%
We note that the structure of the Hamiltonian is two coupled SSH chains,
which is illustrated in Fig. \ref{fig3}.

We consider a simple case with $\sqrt{\Delta _{\alpha }\Delta _{\beta }}=t$.
In this situation the Majorana model reduces to%
\begin{equation}
H_{\mathrm{cp}}=-\frac{i}{2}(t\sum\limits_{j=1}^{N-1}b_{j}a_{j+1}+\sum%
\limits_{j=1}^{N}\mu a_{j}b_{j})+\text{\textrm{H.c.}},  \label{SSH}
\end{equation}%
which has the form of the SSH model. For the case $\left\vert \mu
\right\vert <1,t=1$, there are two combined Majorana fermion operators%
\begin{equation}
f_{+}=\frac{1}{\sqrt{\Omega }}\sum\limits_{j=1}^{N}\mu ^{j-1}a_{j}\text{, }%
f_{-}=\frac{1}{\sqrt{\Omega }}\sum\limits_{j=1}^{N}\mu ^{N-j}b_{j},
\end{equation}%
with $\Omega =\left( \mu ^{N}-1\right) /\left( \mu -1\right) $, which obey%
\begin{equation}
\{f_{+},f_{-}\}=0,\text{ }f_{\pm }^{2}=1.
\end{equation}%
and%
\begin{equation}
\lbrack H_{\mathrm{cp}},f_{\pm }]=0.
\end{equation}%
We can construct a standard fermionic operator\ from Majorana operator $%
f_{\pm }$
\begin{equation}
f_{N}=\frac{1}{2}\left( f_{+}-if_{-}\right) ,
\end{equation}%
which obeys%
\begin{equation}
\{f_{N},f_{N}^{\dag }\}=1,(f_{N})^{2}=0.
\end{equation}

Furthermore, the well-known solution of single-particle SSH chain allows us
to construct a set of standard fermionic operators $\{f_{j}\}$ $(j\in
\lbrack 1,N])$, in which only $f_{N}$\ has an analytical expression. The
explicit expression for other $f_{j}$\ is not necessary for the present
work, since we only concern the zero-energy mode. Accordingly, the
Hamiltonian (\ref{SSH}) can be rewritten as%
\begin{equation}
H_{\mathrm{cp}}=\sum\limits_{j=1}^{N-1}\varepsilon _{j}(f_{j}^{\dagger
}f_{j}-\frac{1}{2})+0\times f_{N}^{\dagger }f_{N},
\end{equation}%
where $\varepsilon _{j}$\ corresponds to the spectrum of single-particle SSH
chain. Then $f_{N}$\ represents the zero mode\textbf{,} since it is a
zero-energy eigen operator, satiafying\textbf{\ }%
\begin{equation}
\left[ f_{N},H_{\mathrm{cp}}\right] =0.
\end{equation}%
\textbf{\ }In contrast, there is no zero mode for $\left\vert \mu
\right\vert \geqslant 1$\textbf{.} Similarly, we can get the same conclusion
for the case with $t=-1=-\sqrt{\Delta _{\alpha }\Delta _{\beta }}$, which
corresponds to another type zero mode illustrated in Fig. \ref{fig3}. Fig. %
\ref{fig4} plots the energy spectrum from Eq. (\ref{majorana}), which
indicates the coexistence of zero modes and non-zero Zak phases,
demonstrating the bulk-edge correspondence.

In order to extend the obtained result to the non-Hermitian variant, we
express $f_{N}$\ by\textbf{\ }$\left( c_{j}^{\dagger },c_{j}\right) $\textbf{%
\ }as%
\begin{equation}
f_{N}=\frac{1}{2\sqrt{\Omega }}\sum\limits_{j=1}^{N}\mu
^{j-1}(c_{j}^{\dagger }+c_{j}-c_{N-j+1}^{\dagger }+c_{N-j+1}),
\end{equation}%
According to the\textbf{\ }equivalence relation between $\mathcal{H}$\ and $%
H_{\mathrm{cp}}$ from Eqs. (\ref{eq1}) and (\ref{eq2}), we can construct a
pair of canonical operators $\left( \overline{\digamma }_{N},\digamma
_{N}\right) $%
\begin{eqnarray}
\overline{\digamma }_{N} &=&\frac{1}{2\sqrt{\Omega }}\sum\limits_{j=1}^{N}%
\mu ^{j-1}\left( \overline{d}_{j}+d_{j}+\overline{d}_{N-j+1}-d_{N-j+1}\right)
\notag \\
&=&\frac{1}{2\sqrt{\Omega }}\sum\limits_{j=1}^{N}\mu ^{j-1}[\sqrt{\frac{%
\Delta _{\beta }}{\Delta _{\alpha }}}\left( c_{j}^{\dagger
}+c_{N-j+1}^{\dagger }\right)  \notag \\
&&+\sqrt{\frac{\Delta _{\alpha }}{\Delta _{\beta }}}\left(
c_{j}-c_{N-j+1}\right) ],
\end{eqnarray}%
and%
\begin{eqnarray}
\digamma _{N} &=&\frac{1}{2\sqrt{\Omega }}\sum\limits_{j=1}^{N}\mu
^{j-1}\left( \overline{d}_{j}+d_{j}-\overline{d}_{N-j+1}+d_{N-j+1}\right)
\notag \\
&=&\frac{1}{2\sqrt{\Omega }}\sum\limits_{j=1}^{N}\mu ^{j-1}[\sqrt{\frac{%
\Delta _{\beta }}{\Delta _{\alpha }}}\left( c_{j}^{\dagger
}-c_{N-j+1}^{\dagger }\right)  \notag \\
&&+\sqrt{\frac{\Delta _{\alpha }}{\Delta _{\beta }}}\left(
c_{j}+c_{N-j+1}\right) ],
\end{eqnarray}%
which obey%
\begin{equation}
\{\digamma _{N},\overline{\digamma }_{N}\}=1
\end{equation}%
and%
\begin{equation}
\lbrack \overline{\digamma }_{N},\mathcal{H}]=\left[ \digamma _{N},\mathcal{H%
}\right] =0.
\end{equation}%
It shows that the existence of the zero mode. And the bulk-edge
correspondence still holds for the original imbalanced Kitaev model in the
unbroken region.

We are interested in the physical picture of the edge modes. We define
edge-mode states as%
\begin{eqnarray}
\left\vert \psi _{\text{L}}\right\rangle &=&\frac{1}{\sqrt{2}}(\overline{%
\digamma }_{N}+\digamma _{N})\left\vert \text{Vac}\right\rangle , \\
\left\vert \psi _{\text{R}}\right\rangle &=&\frac{1}{\sqrt{2}}(\overline{%
\digamma }_{N}-\digamma _{N})\left\vert \text{Vac}\right\rangle ,
\end{eqnarray}%
which are believed to capture the feature of the zero modes. Obviously, we
have%
\begin{eqnarray}
\left\vert \psi _{\text{L}}\right\rangle &=&\sqrt{\frac{\Delta _{\beta }}{%
2\Omega \Delta _{\alpha }}}\sum\limits_{j=1}^{N}\mu ^{j-1}\left\vert
j\right\rangle , \\
\left\vert \psi _{\text{R}}\right\rangle &=&\sqrt{\frac{\Delta _{\beta }}{%
2\Omega \Delta _{\alpha }}}\sum\limits_{j=1}^{N}\mu ^{j-1}\left\vert
N-j+1\right\rangle ,
\end{eqnarray}%
where $\left\vert j\right\rangle =c_{j}^{\dagger }\left\vert \text{Vac}%
\right\rangle $ is a position state. We find that $\left\vert \psi _{\text{L}%
}\right\rangle $\ is the reflection of $\left\vert \psi _{\text{R}%
}\right\rangle $\ about the center at $\left( N-j+1\right) /2$ and both
states are evanescent bound states at two ends. The decay rate is determined
by $\mu $, while the asymmetry ratio $\Delta _{\beta }/\Delta _{\alpha }$
gives an overall factor. We would like to point out that, state $\left\vert
\psi _{\text{L,R}}\right\rangle $\ is not an eigenstate of the Hamiltonian,
but the one related to the Majorana mode.

\section{Summary}

\label{sec_summary}

In summary, we have analyzed a 1D non-Hermitian Kitaev model that exhibits
the similar topological features of a Hermitian one, the bulk-edge
correspondence within the unbroken time-reversal symmetric regions, in which
the Zak phase is defined in the context of biorthonormal inner product and
the edge mode is obtained in the aid of the corresponding Hermitian
counterpart. It indicates that the imbalance of pair creation and
annihilation in some extents, does not destroy the superconducting phases
and their topological features. The bulk-edge correspondence is immune to
the non-Hermitian effect. For extreme imbalance, spontaneously time-reversal
symmetry breaking occurs, corresponding to the emergence of gapless phase
arising from coalescing state. We also examined the profile of the edge
modes in the spinless fermionic representation. It shows that the Majorana
edge modes associate with bound fermionic states at two ends of a Kitaev
chain. This may provide a way to detect the existence of Majorana fermions.

\acknowledgments We acknowledge the support of the CNSF (Grant No. 11374163).

\end{document}